# Perception Lie Paradox: Mathematically Proved Uncertainty about Humans Perception Similarity

Ahmed M. Mahran

Computer and Systems Engineering Department, Faculty of Engineering,
Alexandria University, Alexandria 21544, Egypt

Email: ahmahran@gmail.com

*Abstract* – Agents' judgment depends on perception and previous knowledge. Assuming that previous knowledge depends on perception, we can say that judgment depends on perception. So, if judgment depends on perception, can agents judge that they have the same perception? In few words, this is the addressed paradox through this document. While illustrating on the paradox, it's found that to reach agreement in communication, it's not necessary for parties to have the same perception however the necessity is to have perception correspondence. The attempted solution to this paradox reveals a potential uncertainty in judging the matter thus supporting the skeptical view of the problem. Moreover, relating perception to intelligence, the same uncertainty is inherited by judging the level of intelligence of an agent compared to others not necessarily from the same kind (e.g. machine intelligence compared to human intelligence). Using a proposed simple mathematical model for perception and action, a tool is developed to construct scenarios, and the problem is addressed mathematically such that conclusions are drawn systematically based on mathematically defined properties. When it comes to formalization, philosophical arguments and views become more visible and explicit.

## I. Introduction

Perception is the process of feeling and understanding the environment by ways of sensory data processing. These sensory data are the results of interactions between sensors and outer stimuli [1]. Hence, it could be said that perception is the process of translating the outside world to an inside representation into the perceiver. This directly brings up to mind the other way translation: The translation from the inside representation into the outside world. This is what is called behavior or action. The term "agent" is used to denote an entity that has these translation capabilities, an entity that is able to perceive and act.

Being the only port to the world, perception has been a source of problems and engrossment to philosophers [2] [3]. The problem addressed in this document is not concerned with the quality of the perception process (whether there is illusion, hallucination, insensitivity to simulus changes, or the perciever is not capable of sensing some stimulus ranges, …). The problem is about the possibility of judging the relation between the received percepts of two agents for the same stimulus in terms of similarity.

Although this problem has been addressed before such that many scenarios have been composed and used in philosiphcal arguments [4] [5], the problem here is addressed differently. Using a proposed simple mathematical model for perception and action, a tool is developed to construct scenarios, and the problem is addressed mathematically such that conclusions are drawn systematically based on mathematically defined properties. When it comes to formalization, philosophical arguments and views become more visible and explicit.

## II. A Mathematical Model of Perception-Action

Consider an agent, $A$, that generates a stimulus, $s$, of type $S$ with a function $f_{S,A}(x) = s$ where $x$ is a parameter that the value of the stimulus, $s$, depends on. For example, when agent, $A$, says "Hello", it generates a sound wave corresponding to that "Hello". So, we will call the generated sound wave a *stimulus*, $s$, that corresponds to the word "Hello" which we will call a *parameter* giving it the symbol $x$. The mapping from $x$ to the mathematical representation of the stimulus ($s$) is the function $f_{S,A}$, that function generates a stimulus of type $S$, which is *sound*, that corresponds to a parameter or a meaning $x$, here $x = $ "$Hello$". Figure 1-(a) shows this case.

Also, that agent, $A$, understands a meaning, $x$, by sensing a stimulus, $s$, of type $S$ with a function $f_{S,A}^{-1}(s) = x$ such that $s$ is the received stimulus by the agent from the surrounding world. For example, the agent receives the previously generated sound wave. Then, it finds its meaning by applying the function $f_{S,A}^{-1}$ on that stimulus (sound wave) to find out that $f_{S,A}^{-1}(sound\ wave("Hello")) = "Hello"$. So, it simply hears "Hello". Figure 1-(b) shows this case.

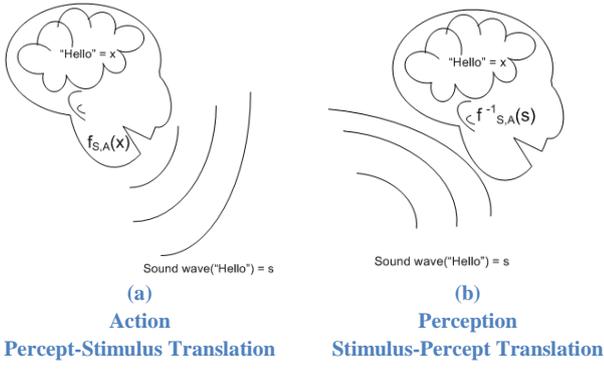

(a) Action
Percept-Stimulus Translation

(b) Perception
Stimulus-Percept Translation

Figure 1 – (a) is an agent understands a meaning "Hello" and expresses it by generating the sound wave of "Hello" while (b) hears the sound wave of "Hello" and understands as "Hello"

Figure 2 shows the graphical symbolic representation of an agent with the previously mentioned functions.

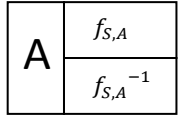

Figure 2 – A graphical symbolic representation of an agent that can generate and perceive a stimulus of type $S$

The line
$$A: f_{S,A}(x)$$
means that: agent $A$ generates a stimulus corresponding to the meaning $x$. In the previous example, to indicate that $A$ said "Hello", we can write

$A: "Hello"$  ; $A$ understands the meaning "Hello"
$A: f_{S,A}("Hello")$  ; $A$ expresses the meaning of "Hello" by saying "Hello"

On the other hand, the line
$$A: f_{S,A}^{-1}(s)$$
means that: agent $A$ understands the meaning of the received stimulus, $s$, as $f_{S,A}^{-1}(s)$ which equals $x$. In the previous example, to say that $A$ hears the previously generated sound of "Hello" and understands it as "Hello", we can write

$A: f_{S,A}^{-1}(sound("Hello"))$  ; $A$ hears the sound of "Hello"
$A: f_{S,A}^{-1}.f_{S,A}("Hello")$  ; $A$ hears the sound of "Hello"
$A: "Hello"$  ; $A$ understands the meaning "Hello"

Agent $A$ is surrounded by a world $W$. We can assume without loss of generality that the world carries the generated stimulus without any modification.

## III. THE PERCEPTION LIE PARADOX

Now, suppose that there are two agents, $A$ and $B$, that can generate and percept sound, having the following dialog. $A$ understands a meaning and expresses it by sound. Then, $B$ hears the sound and understands it. After that, $B$ re-sounds it back to $A$ so as to make sure that $B$ understands the required meaning. The following is the dialog between them.

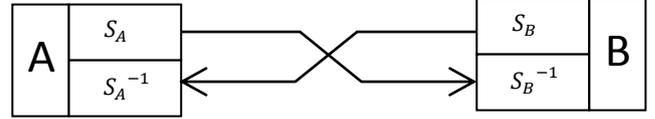

Figure 3 – A dialog between two agents: A and B

$A:$ "Hello"  ; $A$ understands the meaning "Hello"
$A:$ $S_A("Hello")$  ; $A$ says "Hello"
$W:$ $S_A("Hello")$  ; the world $W$ carries the sound of "Hello"
$A:$ $S_A^{-1}.S_A("Hello")$  ; $A$ hears itself and understands that $S_A("Hello")$ is the sound of "Hello"
$A:$ "Hello"  ; $A$ understands the meaning "Hello"
$B:$ $S_B^{-1}.S_A("Hello")$  ; $B$ hears $S_A("Hello")$ and understands the meaning $S_B^{-1}.S_A("Hello")$
$B:$ $S_B.S_B^{-1}.S_A("Hello")$  ; $B$ expresses the meaning $S_B^{-1}.S_A("Hello")$ by saying $S_B^{-1}.S_A("Hello")$ with its sound $S_B.S_B^{-1}.S_A("Hello")$
$B:$ $S_A("Hello")$  ; $B$ tries to generate the same sound it heard before, $S_A("Hello")$, so $S_B$ and $S_B^{-1}$ cancel each other
$W:$ $S_A("Hello")$  ; $W$ carries the sound generated by agent $B$, $S_A("Hello")$
$A:$ $S_A^{-1}.S_A("Hello")$  ; $A$ hears $S_A("Hello")$ and figures out that it is equivalent or like its sound
$A:$ "Hello"  ; $A$ understands the sound $S_A("Hello")$ is the sound of "Hello", so $A$ understands the meaning "Hello"

The following table summarizes what both agents have understood:

| Agent $A$ | Agent $B$ |
| --- | --- |
| "Hello"= $x$ | $S_B^{-1}.S_A("Hello") = y$ |

It could be noticed that agent $B$ understands a different meaning than what agent $A$ understands, however agent $B$ succeeded to make agent $A$ understand what matches $A$'s thoughts or way of thinking! That is a confusing result; both agents think that they understand the same meaning although they may not (or do not)!

Agent $A$ thinks that agent $B$ understands the meaning of the sound as $x$ (the same way $A$ understands it). On the other hand,

$B$ thinks that $A$ understands the meaning of the sound as $y$ (the same way $B$ understands it). They both think that they understand meanings the same way. They both think that they have the same perception of the surrounding world so they think that that's why they can understand each other. However, we can see that they both have different perception as well as they can understand each other well. For both agents to have the same perception, $S_A^{-1}$ should be equal to $S_B^{-1}$. The question is: Can they (either $A$ or $B$) tell that $S_A^{-1} = S_B^{-1}$?

*Agents' judgment depends on perception and previous knowledge. Assuming that previous knowledge depends on perception, we can say that judgment depends on perception without the need to mention previous knowledge. So, if judgment depends on perception, can agents judge that they have the same perception? Can agents be sure that their judgment about perception is not deviated by their perception?*

Consider agents that can interact with five different stimuli: sound ($S$), light ($L$), touch ($C$), taste ($T$), smell ($M$). Can they tell that all $S$'s are equivalent ($S_A^{-1} = S_B^{-1} = S_C^{-1} = S_D^{-1} = \cdots$), all $L$'s are equivalent, all $C$'s are equivalent, all $T$'s are equivalent, and all $M$'s are equivalent? Can they tell that they have the same perception of the world in a way that is not affected by their perception?

*So, the question is NOT "Do humans have the same perception?" but the question is "Can humans judge that they have the same perception?"*

Being agnostic about this question imposes the 50% probability that we, humans, might have different perception of the world around us. It is somehow shocking to think so! One might see that tomatoes, that you know, are red, the red you know, while another might see them blue, the same as your blue, but they cannot figure out that they are perceiving these different meanings. The one that knows that tomatoes are red says to the other one "Tomatoes are red". Then, the other one that knows that tomatoes are blue hears the one with red tomatoes as if saying "Tomatoes are blue" so he knows that he is talking right. So, the one with the blue tomatoes says to the other one "You are right; tomatoes are blue". Then comes the turn of the one with red tomatoes to hear "blue" as if hearing "red" and life continues.

Having different perception of the world, means that there are different worlds! As long as we interpret stimuli differently, we live in different worlds! In my world, I might interpret someone's actions towards me as if being nice so I act nicely while he thinks, in his own world, that he is being rude and I am coping up with him by being rude although I'm being nice in my world, and life still continues! You might describe for me the shape of a circle while I understand your description as a description for a square and when I re-describe the description I understood as a square's description you understand it as a circle's description! So you think I've understood your meaning and, fortunately, life keeps running!

In my world, I might see people with three legs and walking upside down but in your world people might have ten legs and no arms. However, we can still communicate without noticing any difference or anything weird in the other's world. It might be true that Bohr, Planck, Heisenberg, Einstein, Newton and others are those men who had revealed the mystery of your physical world and Muhammad, Jesus, Moses and others are those men who had revealed the mystery of your metaphysical world.

Those sentences might look very weird and so might be the concept of, literally, completely different perceptions which leads to, literally, completely different realities. This is different than saying that: "Real" has different interpretations to some extent. As in the later, if two humans are in that case, they are there because each one did not see/sense what the other had seen/sensed. However, when one is moved to the other's place and senses the same experience as of the other's, the moved one will conclude a nearer interpretation to his opponent, if not the same, and will be understanding the two different interpretations. Or both of them could continue arguing each other and still it's possible for them to reach a common understanding. In some point of time, they will catch the difference.

This is not the case on the other hand. Both are conceived that they understand the same thing the same way, which is not the truth, and they have no mean to feel the complete difference.

IV. PERCEPTION CORRESPONDENCE

*It could be concluded that, an agreement about the understanding of a concept could be reached in communication, regardless from the similarity of the forms of understanding of that concept among communication parties, as long as (i) the communicated stimuli are the same among communication parties and (ii) those stimuli are always triggered by the same form of understanding and always trigger the same form per communication party.*

Mathematically, this means that for all stimuli, there has to be one-to-one correspondence from the form of understanding of one agent to the other. So, for agents $A$ and $B$ to reach an agreement about a concept, it is not necessary that their forms of understanding to be the same ($S_A^{-1} = S_B^{-1}$) however the necessity is to have a one-to-one correspondence function:

$$C: S_B^{-1} \to S_A^{-1} \text{ (i.e. } C(S_B^{-1}) = S_A^{-1})$$

So, if it is possible to find such a function, it could be judged that $A$ and $B$ can reach agreement or they have *perception correspondence*.

Relating perception to intelligence [6] [7], the fact that two agents have perception correspondence contributes to the judgment whether or not they are on the same level of intelligence. If agents $A$ and $B$ understand things in the same way or more generally in a one-to-one correspondence manner, then most probably they do have the same level of intelligence.

## V. A SOLUTION ATTEMPT TO THE PARADOX

There might be a scenario that tells whether $S_A^{-1} = S_B^{-1}$ or not. Increasing complexity of dialogs by adding more agents with more sensors and actuators won't help as long as the couple of a perception function and its inverse cancel each other. However, assumptions that might help could be like making the surrounding world convert one stimulus of any type to another of any other type (we can watch the effect of sound on dust) or give that ability to agents.

Imagine that we have a special kind of agent. An agent that has a special perception function that can percept other agents' percepts. A thinkable scenario is that, the special agent to observe two normal agents perceiving the same stimulus. Then he can conclude whether the observed percepts of both agents are similar or not.

Let's call the special agent a "Judge", $J$, and the special perception function of the judge "Observation", $O_J$. This scenario is illustrated by Figure 4.

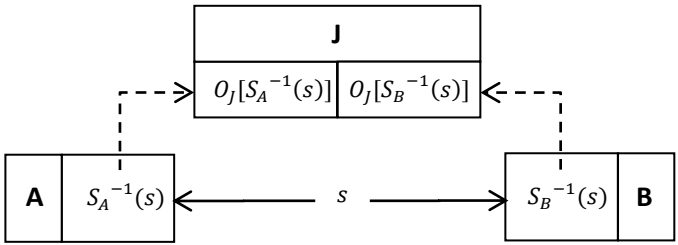

Figure 4 – The judge agent observing percepts of normal agents

The possible observations of the judge are:

1. $O_J[S_A^{-1}(s)] \neq O_J[S_B^{-1}(s)]$. In that case, the judge can conclude that $S_A^{-1}(s) \neq S_B^{-1}(s)$
2. $O_J[S_A^{-1}(s)] = O_J[S_B^{-1}(s)]$. In that case, there are two conclusions:
   a. Either, $S_A^{-1}(s) = S_B^{-1}(s)$
   b. Or, $O_J$ is not one-to-one function which means that for some values of $s$, $S_A^{-1}(s) = S_B^{-1}(s)$ and for other values of $s$, $S_A^{-1}(s) \neq S_B^{-1}(s)$

*In conclusion, even that conceptual judge can only be sure that agents have different perception and cannot be sure that agents have the same perception. In other words, if this judge is asked whether agents have the same perception or not, he can say "No" with a complete certainty of 100% but cannot say "Yes" with such certainty.*

## VI. CERTAINTY ABOUT PERCEPTION CORRESPONDENCE

Besides judging whether both agents have the same perception or not, this conceptual judge might make further judgment about the perception correspondence between both agents. Looking again at the possible observations of the judge:

1. $O_J[S_A^{-1}(s)] \neq O_J[S_B^{-1}(s)]$ and the judge concludes that $S_A^{-1}(s) \neq S_B^{-1}(s)$. In that case:
   a. If it is possible for the judge to find the perception correspondence function
      $C_{JBA}: O_J[S_B^{-1}(s)] \to O_J[S_A^{-1}(s)]$ (i.e.
      $C_{JBA}(O_J[S_B^{-1}(s)]) = O_J[S_A^{-1}(s)]$), then the judge can conclude that $A$ and $B$ can reach agreement, they have perception correspondence, or they might be at the same level of intelligence.
   b. Or, if it is not possible for the judge to find such function, then he can conclude the converse.
2. $O_J[S_A^{-1}(s)] = O_J[S_B^{-1}(s)]$ and either $S_A^{-1}(s) = S_B^{-1}(s)$ or $S_A^{-1}(s) \neq S_B^{-1}(s)$ could be true. In that case, the correspondence function is easily found as
   $C_{JBA}(O_J[S_B^{-1}(s)]) = O_J[S_A^{-1}(s)] = O_J[S_B^{-1}(s)]$
   however $J$ might be falsely judging that $A$ and $B$ can reach agreement, they have perception correspondence, or they might be at the same level of intelligence.

*Another conclusion is that, the degree of certainty of the conceptual judge judgment about whether agents have perception correspondence or not (might be at the same level of intelligence or not) is inherited from the certainty of that conceptual judge judgment about whether the same agents have the same perception or not.*

## VII. CERTAINTY ABOUT WIDE PERCEPTION CORRESPONDENCE

Introducing a new agent, $M$, from different type than of $A$ and $B$ ($M$ is an intelligent machine), $J$ will have a different observation function for that kind, $O_{JM}$. Let's consider the scenario illustrated in Figure 5:

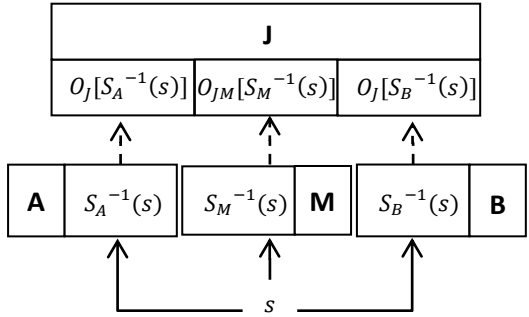

Figure 5 – The judge agent observing percepts of normal agents $A$ and $B$ and an intelligent machine $M$

If it is possible for the judge to find the following correspondence functions:

$$C_{J_{MA}}(O_{JM}[S_M^{-1}(s)]) = O_J[S_A^{-1}(s)]$$

$$C_{J_{MB}}(O_{JM}[S_M^{-1}(s)]) = O_J[S_B^{-1}(s)]$$

1. Either the judge can conclude that $M$ has a wide/general perception correspondence (or a wide/general intelligence) if $O_J[S_A^{-1}(s)] \neq O_J[S_B^{-1}(s)]$.
2. Or the judge can falsely conclude that $M$ has a wide perception correspondence (or a wide/general intelligence) if $O_J[S_A^{-1}(s)] = O_J[S_B^{-1}(s)]$.

*As a conclusion, the degree of certainty about whether an agent (e.g. an intelligent machine) has a wide/general perception correspondence (wide/general intelligence) or not compared to some kind of agents is inherited from the certainty about whether these agents have the same perception or not.*

So, solving the perception paradox gives useful hints about the intelligence of an agent compared to others not necessarily from the same kind.

## VIII. CONCLUSION

Perception has been a source of problems and engrossment to philosophers. Many philosophical problems had been raised and many views had been proposed. The proved uncertainty about judging perception similarity among agents supports the philosophical skeptical view of the world and highlights one limitation of the human mind. This uncertainty is automatically extended to everything related in some way or dependent on perception. Also, the mathematical formulation of the problem gives clear and explicit explanation and provides access to direct and systematic conclusions.